\documentclass[9pt,twocolumn,twoside]{pnas-new}
\usepackage{braket} 
\templatetype{pnasresearcharticle} 
\graphicspath{{pics/}}

\title{Charged String Tensor Networks}

\author[a,b,1]{Jacob Biamonte}

\affil[a]{Quantum Complexity Science Initiative, Department of Physics, University of Malta, MSD 2080}
\affil[b]{Institute for Quantum Computing, University of Waterloo, Waterloo, ON N2L 3G1, Canada}

\leadauthor{Biamonte} 

\significancestatement{Liu, Wozniakowski and Jaffe developed a graphical language to reason about quantum information using charged strings in three manifolds and discovered several remarkable braidings for universal quantum computing gates \cite{2016arXiv161202630L}. }

\authorcontributions{JB wrote this paper.}
\authordeclaration{The author declares no conflicts of interest.}
\correspondingauthor{\textsuperscript{1}\url{jacob.biamonte@qubit.org} $|$ \url{www.QuamPlexity.org}}


\begin{abstract}
Tensor network methods provide an intuitive graphical language to describe quantum states, channels, open quantum systems and a class of numerical approximation methods that efficiently simulate certain many-body states in one spatial dimension.  There are two fundamental types of tensor networks in wide use today.  The most common is similar to quantum circuits.  The second is the braided class of tensor networks, used in topological quantum computing.  Recently a third class of tensor networks was discovered by Jaffe, Liu and Wozniakowski---the JLW-model---notably, the wires carry charge excitations.  The rules in which network components can be moved, merged and manipulated in a graphical form of reasoning take an elegant form.  For instance the relative charge locations on wires carries precise meaning and changing the ordering modifies a connected network specifically by a complex number.  The type of isotopy discovered in the topological JLW-model provides an alternative means to reason about quantum information, computation and protocols.  Here we recall the tensor-network building blocks used in a controlled-NOT gate.  Some open problems related to the JLW-model are given.

\end{abstract}

\dates{This manuscript was compiled on \today}
\doi{\url{www.pnas.org/cgi/doi/10.1073/pnas.XXXXXXXXXX}}

\begin{document}

\verticaladjustment{-2pt}

\maketitle
\thispagestyle{firststyle}
\ifthenelse{\boolean{shortarticle}}{\ifthenelse{\boolean{singlecolumn}}{\abscontentformatted}{\abscontent}}{}

\dropcap{T}ensor networks in physics can be traced back to a 1971 paper by Penrose \cite{Pe71}.  Such network diagrams appear in digital circuit theory, and form the foundations of quantum computing---starting with the work of Feynman and others in the 1980's \cite{Feynman1986} and further extended by Deutsch in his `quantum computational network model' \cite{Deutsch73}.   Building on a series of results \cite{Jaffe2016, 2016arXiv160500127J, 2016arXiv161106447J}  Liu, Wozniakowski and Jaffe have recently developed a topological variant of tensor networks which among other results, lead to their discovery of an elegant charged string braiding for the controlled-NOT gate (a.k.a.~the Feynman gate) \cite{2016arXiv161202630L}.  

Category theory is a branch of mathematics well suited to describe a wide range of networks \cite{2009arXiv0908.2469B}.  Quantum circuits were first given a `categorical model' in pioneering work by Lafont in 2003 \cite{Lafont03towardsan} and dagger compact closed categories \cite{BD95}, also called Baez-Dolan $\dagger$-categories, were first derived to describe both standard quantum theory as well as classes of topological quantum field theories in seminal work published in 1995 \cite{BD95}.  (see \cite{2009arXiv0908.2469B} for a well written review of categorical quantum mechanics.)  Liu, Wozniakowski and Jaffe formulated their topological model, in part, using category theory  \cite{2016arXiv161202630L}.  For practical purposes, the graphical language turns out to be mathematically equivalent to the categorical formulation.  So one can work with the diagrams.  But for those worried about the formalism, it's been fully worked out and applied to quantum physics for a few decades now, pioneered in the seminal work of Baez and Dolan \cite{BD95} and connected to quantum circuits by Lafont \cite{Lafont03towardsan}. 


\begin{figure}[h]
\includegraphics[width=0.48\textwidth]{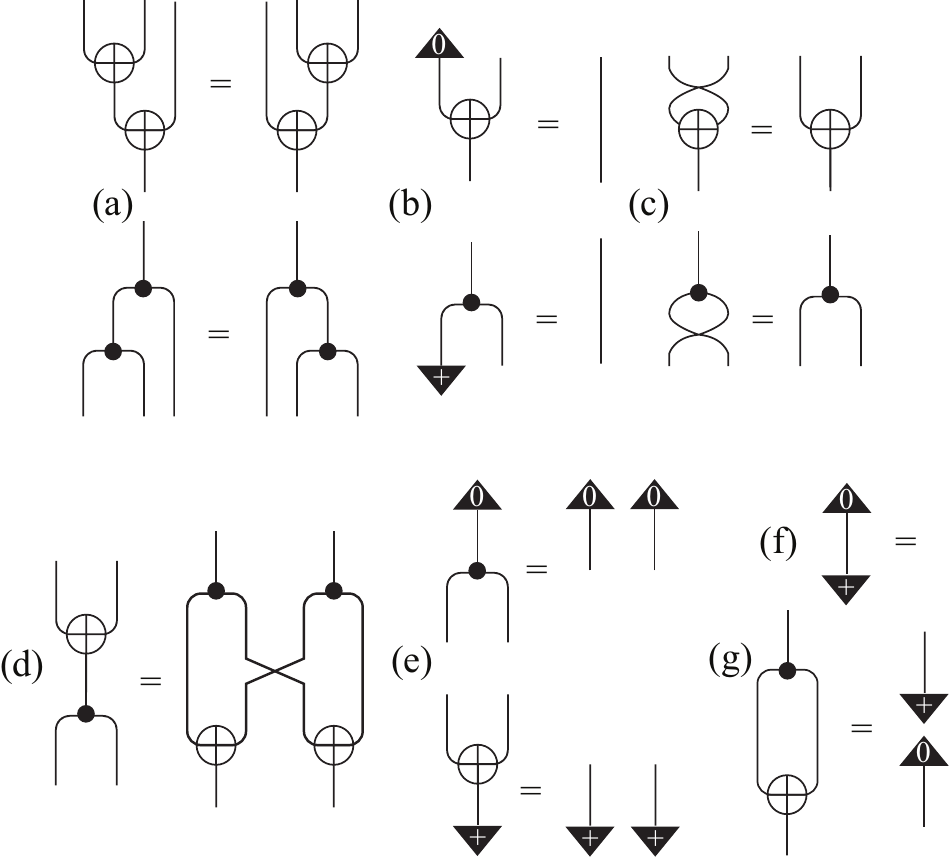}
\caption{Lafont's 2003 categorical model of quantum circuits included the bialgebra (d) and Hopf (g) relations between the building blocks needed to form a controlled-NOT gate.  The practical utility of using Baez-Dolan $\dagger$-categories \cite{BD95} to describe quantum circuits \cite{Lafont03towardsan, 2016arXiv161106447J}, is that category theory provides a graphical language which fully dictates the types of admissible relationships and transformations to reason about the interaction of network components.  Redrawn from \cite{Lafont03towardsan}.  [(a) associativity; (b) gate unit laws; (c) symmetry; (e) copy laws; (f) unit scalar given as a blank on the page.]  }\label{fig:F2-presentation}
\end{figure}

Lafont's algebraic theory of logic gates \cite{Lafont03towardsan} was cast into the setting of tensor network states---as used at the crossroads of condensed matter theory and quantum computation---by myself with several colleagues \cite{2011AIPA....1d2172B, 2011JPhA...44x5304B, 2013JPhA...46U5301B}.  We adapted these tools and discovered efficient tensor network descriptions of finite Abelian lattice gauge theories \cite{2012JPhA...45a5309D}.  These tools also lead to the discovery of a wide class of efficiently contractable tensor networks, representing counting problems \cite{2015JSP...160.1389B}.  


The Feynman gate is a (if not the) central building block for quantum information processing tasks.  In the braided tensor network model, the braiding sequences to represent a Feynman gate are painfully complicated---resembling perhaps a musical score---and hence salient intuitive features are lacking.   The known tensor network frameworks differ strikingly from the topological underpinnings in \cite{2016arXiv161202630L}.  Constructing the Feynman gate, as a concatenation of building blocks is however, similar.  Let's recall the building blocks used in the tensor network construction of the Feynman gate. 

\subsection*{Tensor network building blocks}
A universal model of computation can be expressed in terms of networks (i.e.~circuits built from gates).  The first gate to consider copies binary inputs ($0$ and $1$) like this 
\begin{subequations}
\begin{align}
        0 &\rightarrow 0,0\label{eqn:copy}\\
        1 &\rightarrow 1,1\label{eqn:copy2}
\end{align}
\end{subequations}
In the diagrammatic tensor network language, the copy-gate is 
\begin{center}
\includegraphics[]{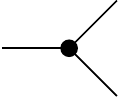}
\end{center} 
and graphically, equation \ref{eqn:copy} and \ref{eqn:copy2} become 
\begin{center}
\includegraphics{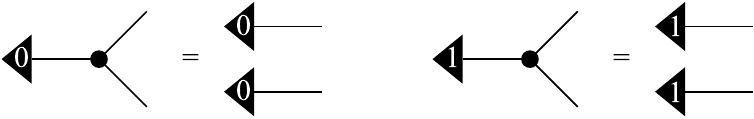}
\end{center} 
The next gate preforms the exclusive OR operation (XOR).  Given two binary inputs (say $a$ and $b$), the output ($a\oplus b = a + b - 2 ab$) is $1$ iff exactly a single input is $1$ (that is, addition modulo 2).  The gate is drawn as 
\begin{center}
\includegraphics{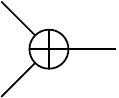}
\end{center} 
The XOR gate allows one to realize any linear Boolean function.  Let $f$ be a function from $n$-long bit strings $x_1 x_2\dots x_n$ to single bits $\in \{0,1\}$. Then $f(x_1,x_2,\dots,x_n)$ is linear over $\oplus$ if it can be written as 
\begin{equation}\label{eqn:linear}
f = c_1 x_1 \oplus c_2 x_2 \oplus \cdots \oplus c_{n-1} x_{n-1} \oplus c_n x_n
\end{equation}
where ${\bf c} := (c_1, c_2, \dots ,c_{n-1}, c_n)$ is any $n$-long Boolean string.  Hence, there are $2^n$ linear Boolean functions and note that negation is not allowed.  When negation is allowed a constant $c_0\oplus f$ is added (mod 2) to equation \ref{eqn:linear} and for $c_0=1$, the function is called affine. In other words, negation is equivalent to allowing constant $1$ as 
\begin{equation}\label{eqn:constants}
\includegraphics{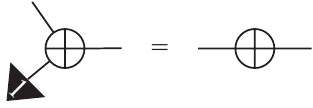}
\end{equation}
which sends Boolean variable $x$ to $1-x$.  Using the polarity representation of $f$, 
\begin{equation}
\hat f ({\bf x}) = (-1)^{f({\bf x})}
\end{equation}
we note that linear Boolean functions index the columns of the $n$-fold tensor product of $2\times 2$ Hadamard matrices (that is, $H^{\otimes n}$ where the $i$--$j$th entry of each $2\times 2$ is $\sqrt{2}H_{ij} := (-1)^{i\cdot j}$). Importantly, 
\begin{equation}\label{eqn:xorH}
\includegraphics{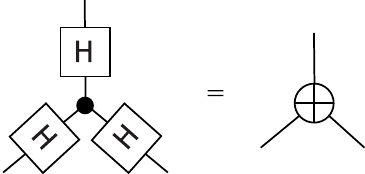}
\end{equation} 
up to isometry as there could be an omitted scale factor depending on conventions.  By equation \ref{eqn:xorH} one can think of XOR as being a copy operation in another basis.  We send binary $0$ to $\ket{0}:=(1,0)^\top$ and $1$ to $\ket{1}:=(0,1)^\top$ where $\top$ is transpose. Then XOR acts as a copy operation: 
\begin{subequations}
\begin{align}
        \ket{+} &\rightarrow \ket{+,+}\label{eqn:xcopy}\\
        \ket{-} &\rightarrow \ket{-,-}\label{eqn:xcopy2}
\end{align}
\end{subequations}
using $H^2 = {\bf 1}$, $\ket{+} := H\ket{0}$ and $\ket{-} := H\ket{1}$.  

Concatenating the copy- and XOR gates \cite{2011JPhA...44x5304B} yields the logically reversible Feynman gate 
\begin{equation}\label{fig:feynman}
\includegraphics{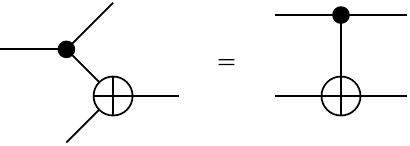}
\end{equation} 

A simplistic methodology to connect quantum circuits with indexed tensor networks starts with the definition of two tensors, in terms of components. 
\begin{center}
\includegraphics{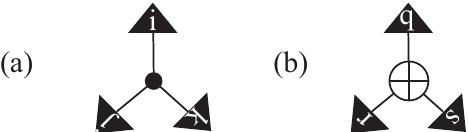}
\end{center} 
In (a) we have 
\begin{equation}
\delta^i_{~jk} = 1 - (i + j + k)+ i j + i k + j k 
\end{equation} 
where the indicies ($i$, $j$ and $k$) take values $\in \{0,1\}$. In other words, the following contractions evaluate to unity. 
\begin{center}
\includegraphics{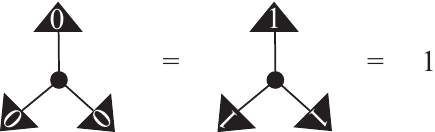}
\end{center} 
Likewise for (b) we have 
\begin{equation} 
\oplus^q_{~r s} = 1-(q+r+s)+2(qr+qs+sr)-4qrs
\end{equation} 
where the following contractions evaluate to unity (the XOR tensor is fully symmetric, hence the three rightmost contractions are identical by wire permutation). 
\begin{center}
\includegraphics{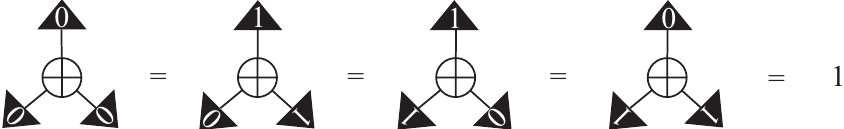}
\end{center} 
Then the Feynman gate (CN) is given as the following tensor contraction\footnote{Equation \ref{eqn:cnot} expands to $1 - (i + j + q + r) + i j + i q + j q + i r + j r + 2 (q r - i q r - j q r)$} 
\begin{equation}\label{eqn:cnot}
\sum_m\delta^{ij}_{~~m}\oplus_{~qr}^{m} = {\text {CN}}^{ij}_{qr}
\end{equation}
where we raised an index on $\delta$.  
All quantum circuits can be broken into their building blocks and thought of as indexed tensor contractions in this way.

\subsection*{Reversible logic}
A reversible computer is built using gates that implement bijective functions.  Quantum gates are unitary: hence reversible classical gates are a subclass.  Let us recall the critical implication of reversible logic.  

We will consider $n$-long bit strings in lexicographic order indexed by natural numbers $i$. So $y_0 = 00\cdots 0$, $y_2 = 00\cdots 10$ etc.~We will further consider inputs as being uniformly distributed over the $y$'s and define the change in Shannon's entropy between a circuits input and output (implementing $g$) as 
\begin{equation}\label{eqn:deltaS}
\Delta S := \sum_i P\{g(y_i)\} \ln_2 P\{g(y_i)\} -  \sum_i P(y_i) \ln_2 P(y_i) 
\end{equation}
where the probability $P\{y_i\} = 2^{-n}, \forall i$ for the uniform distribution.  Equation \ref{eqn:deltaS} vanishes identically iff $g$ is a reversible function.  

For $g$ non-reversible (a.k.a.~a non-injective surjective function), there exists at least one pair $y_i$, $y_j$ such that $g(y_j)=g(y_i)$ and hence, information is lost as the input can not be uniquely recovered from the output (so the Shannon entropy of the output distribution is strictly $<n$) and hence, equation \ref{eqn:deltaS} is non-vanishing.  The vanishing of Eq.~\ref{eqn:deltaS} is a central implication of reversible computation, and provides an abstract argument related to Landauer's principle. 

Universal classical computation can be realized with reversible logic gates.  However, using the Feynman gate is not enough since it only can be used to implement linear functions.  An additional reversible gate must be added, such as the Toffoli or Fredkin gate(s).  (see open problem 1).  However, the Feynman gate can provide universal quantum computation provided specific one-qubit gates are included.  Such an approach was taken by Liu, Wozniakowski and Jaffe in \cite{2016arXiv161202630L}. 


\subsection*{Stabilizer tensor networks} Stabilizer circuits use gates from the normalizer of the $n$ qubit Pauli group---generated by the Clifford gates \cite{DBLP:journals/corr/Selinger13}.  The gates include: (i) the single qubit Pauli gates $X$, $Y$ and $Z$; (ii) the Feynman gate; (iii) the Hadamard gate; (iv) the phase gate $S=\ket{0}\bra{0}+i\ket{1}\bra{1}$.  

We will illustrate that: (a)  a vector $\ket{t}:= \ket{0}+i\ket{1}$, (b) the Hadamard gate and (c) the XOR- and copy tensors and (d) a covector $\bra{+}:= \bra{0}+\bra{1}$ can be contracted to simulate any stabilizer quantum circuit.  We will establish this by recovering the Clifford gates (i-iv).  

By linearity, the copy tensor induces a product between vector pairs, producing a third vector where the coefficients of the input vector pair are multiplied.  This allows us to recover from (a) the family $\ket{t^k}:= \ket{0}+i^k\ket{1}$ for integer $k$ as, for instance, 
\begin{equation}
\includegraphics{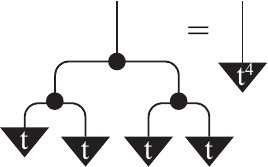}
\end{equation} 
which recovers the vector $\sqrt{2}\ket{+}$ from equation \ref{eqn:xcopy}.  

Then from (d) we can recover a cup (or cap) allowing one to raise/lower indicies, and importantly, \ref{eqn:tcup}.~(b) illustrates that 
\begin{equation}\label{eqn:tcup}
\includegraphics{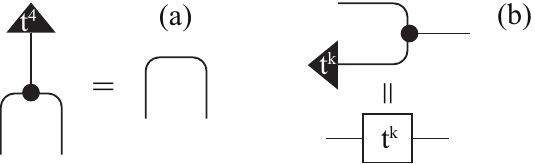}
\end{equation} 
$\ket{t^k}:= \ket{0}+i^k\ket{1}$ lifts to a unitary operator where the Clifford gate $S$ is recovered for $k=1$---thereby establishing (iv).  For $k=2$ we recover the standard Pauli $Z$ matrix, then $HZH= X$ and $SXS^3 = Y$.  So we recover the Pauli gates (i).  The Feynman gate was constructed in \ref{fig:feynman} establishing (ii) and the Hadamard gate (iii) was assumed.

\subsection*{Frontiers in the JLW-model}
We will conclude by stating a few open challenges that relate to the expressiveness and applications opened up by the JLW-model.  There are of course many avenues that can be explored.  The challenges listed here are related to quantum computation (applying the JLW-model to develop new quantum protocols has been considered in \cite{2016arXiv161106447J}).

\subsection*{Open Problem 1}  (Universal reversible logic in the JLW-model).  In principle, the JLW-model has the capability to simulate universal quantum computation.  There are hence indirect methods to create a Toffoli-gate (which implements a reversible variant of the logical AND operation).  However, a direct mapping to universal reversible computing (without having to necessarily create these operations using gates that are outside of the language) would be highly desirable.  Several 3-body gates exist---i.e.~the Fredkin gate---and determining an elegant braiding relation for one of these gates would be an interesting advancement that could (i) pave the way for a topological understanding of reversible logic and (ii) moreover be used in quantum algorithms that require such gates in their oracles.  

\subsection*{Open Problem 2}  (Gottesman--Knill theorem by confluent rewrites in the JLW-model).  
The Gottesman--Knill theorem \cite{stabs} proves that the Clifford gates can be efficiently simulated classically.  
The Clifford operations take a particularly elegant form inside the JLW-model.  
Formulating a graphical proof of the Gottesman--Knill theorem would (i) further demonstrate the expressiveness of the language (ii) bridge the JLW-language with stabilizer theory (which provides an alternative means to define resource states in terms of the stabilizer generators) and (iii) lead possibly to extensions of the theorem to a more general setting of parafermions and qudits and (iv) set the stage for future work on quantum algorithms.

\subsection*{Open Problem 3}  (Quantum circuit simulation of parafermions via the JLW-model).  The main focus to date has been on mapping circuit based protocols to the JLW-model and then searching for new applications and insights inside the topological setting that the language provides.  This mapping can be inverted, providing a new means to simulate a wide class of charge excitations on strings.  An open question is to formalize this mapping on dual-grounds.  Firstly, extending existing simulation methods for e.g.~fermions \cite{2011MolPh.109..735W} would produce gate sequences emulating Hamiltonians that exhibit parafermionic excitations.  Secondly, these excitations can be simulated using gate operations by simulating the JLW-model directly.  Quantifying the overheads and determining the complexity of these operations can lead to a library to simulate a wide array of particles with a quantum computer.

\acknow{I acknowledge the Foundational Questions Institute (FQXi) for financial support. Diagrams are courtesy of Lusa Zheglova (illustrator).}  

\showacknow{} 


\bibliography{pnas-sample}



\end{document}